\documentstyle[11pt,aaspp4,epsf]{article}

\begin{document}
\title{Evidence for ``Propeller'' Effects In X-ray Pulsars GX~1+4 And GRO~J1744-28}
\author{Wei Cui\altaffilmark{1}}
\altaffiltext{1}{Room 37-571, Center for Space Research, Massachusetts Institute of Technology, Cambridge, MA 02139}
\authoremail{cui@space.mit.edu}
\slugcomment{Submitted to the {\it Astrophysical Journal Letters} on 2/25/97; revised on 3/16/97.}
\lefthead{Wei Cui}
\righthead{Propeller Effects in X-ray Pulsars}

\begin{abstract}
We present observational evidence for ``propeller'' effects in two X--ray 
pulsars, GX~1+4 and GRO~J1744--28. Both sources were monitored regularly 
by the {\it Rossi X--ray Timing Explorer} (RXTE) throughout a decaying 
period in the X--ray brightness. Quite remarkably, strong X--ray pulsation 
became unmeasurable when total X--ray flux had dropped below a certain 
threshold. Such a phenomenon is a clear indication of the propeller effects
which take place when pulsar magnetosphere grows beyond the co-rotation
radius as a result of the decrease in mass accretion rate and centrifugal
force prevents accreting matter from reaching the magnetic poles. The entire 
process should simply reverse as the accretion rate increases. Indeed, 
steady X--ray pulsation was reestablished as the sources emerged from the 
non-pulsating faint state. These data allow us to directly derive the surface 
polar magnetic field strength for both pulsars: $3.1\times 10^{13}$~G for 
GX~1+4 and $2.4\times 10^{11}$~G for GRO~J1744-28. The results are likely to 
be accurate to within a factor of 2, with the total uncertainty dominated 
by the uncertainty
in estimating the distances to the sources. Possible mechanisms for the 
persistent emission observed in the faint state are discussed in light of
the extreme magnetic properties of the sources.
\end{abstract}


\section{Introduction}
In accreting X--ray pulsars, the strong magnetic field disrupts accretion 
flow at several hundred neutron star radii and funnels material onto the 
magnetic poles 
(e.g., \cite{pringle1972}; \cite{lamb1973}). X--ray emission mostly comes from 
the ``hot spots'' formed around one or both poles. Such emission is beamed 
(\cite{basko1976}) and thus produces X--ray pulsation by periodically 
passing through the line of sight as the neutron star rotates, if the 
magnetic and rotation axes of the neutron star are misaligned.

The strength of the magnetic field in X--ray pulsars can be illustrated by the 
size of the magnetosphere, which co-rotates with the neutron star. In
the process of mass accretion, the ram pressure of the flow is exerted on the
magnetosphere and is balanced by the magnetic pressure. Therefore, the 
magnetospheric radius is determined not only by field strength but also 
by mass accretion rate. In a bright state, the accretion rate is high, so 
the magnetosphere is usually small compared to the co-rotation radius at 
which the angular velocity of Keplerian motion is equal to that of the 
neutron star. Material is continuously channeled to magnetic poles, so the 
X--ray emission from ``hot spots'' pulsates persistently. As the accretion 
rate decreases, the ram pressure decreases, and thus the magnetosphere 
expands. As the magnetosphere grows beyond the co-rotation radius, 
centrifugal force prevents material from entering magnetosphere, and thus 
accretion onto magnetic 
poles ceases (\cite{pringle1972}; \cite{lamb1973}; \cite{illarionov1975}). 
Consequently, no X--ray pulsation is expected. This is commonly known as 
the ``propeller'' effect, because accreting matter is likely to be ejected in 
the presence of a strong magnetic field.

Although the propeller effects were predicted theoretically in the 1970's, 
there has been no direct observational evidence for them. If the ``standard''
pulsar theory is on the right track, such effects should be observable. A 
positive detection would not only confirm our understanding of X--ray 
pulsars but would also
allow a direct determination of magnetic field strength in these systems. In 
this {\it Letter}, we report the discovery of such effects in two X--ray 
pulsars, GX~1+4 and GRO~J1744-28, based on the RXTE observations.

GX~1+4 is a 2 minute pulsar with a low-mass M giant companion. It is highly
variable in X-rays. The observed X-ray flux can vary by 2 orders of magnitude
on a time scale of months. It has the hardest X-ray spectrum among known X-ray
pulsars. Secular spin-ups and spin-downs have been observed at nearly equal
rates (see \cite{chakrabarty1997b}, and references therein). It has been,
therefore, postulated that GX~1+4 is near spin equilibrium. Then, according to
the ``standard'' model (\cite{ghosh1979}), the observed spin-down rate would 
require a surface magnetic field of a few $\times 10^{13}$ G. Hints for 
such an unusually strong magnetic field are also provided by the observed hard
X--ray spectrum.

GRO~J1744-28 is a transient X-ray pulsar with a period of 0.467 s and 
is thought to have a low-mass companion. It is the only known X--ray 
pulsar that 
also produces X--ray bursts and only the second source, after the Rapid 
Burster, that displays type~II bursts (\cite{lewin1996}). The Rapid burster 
apparently only has a weak magnetic field. By analogy, GRO~J1744-28 may also 
be a weakly magnetized system, but a strong enough field to be an X--ray 
pulsar. 

\section{Observations and Analysis}
Both GX~1+4 and GRO~J1744-28 were monitored regularly by {\it RXTE} in 1996. 
For this study, we only use data from the {\it Proportional Counter Array} 
(PCA). The PCA consists of five nearly identical large area proportional 
counter units (PCUs). It has a total collecting area of about 6500 $cm^2$, 
but for a specific observation some PCUs (up to two) may be turned off for 
safety reasons. It covers an energy range from 2-60 keV with a moderate 
energy resolution ($\sim 18\%$ at 6 keV). Mechanical collimators are used to 
limit the field of view to a $1^{\circ}$ FWHM. 

\subsection{X--ray Light Curves}
Starting on 1997 February 17, GX~1+4 was observed roughly monthly by the
PCA for $\sim$10 ks for each observation. Initially, it was in a relatively 
bright state, $\sim$100 mCrab, with a large pulse fraction ($\sim$60\%). The 
X--ray brightness decayed steadily since that measurement, with the pulse 
fraction remaining roughly constant. By 1996 September 25, it appeared only 
as a $\sim$2 mCrab source. Importantly, the X--ray pulsation was not measured 
by using standard techniques such as fast Fourier transformation and 
epoch-folding (using a detection threshold of
$3\sigma$). For a better coverage of the source
during this interesting period, we chose to observe it weekly for $\sim$5 ks 
each. GX~1+4 remained in this faint state (\cite{chakrabarty1996}; 
\cite{cui1996}) until around 1996 November 29, when it brightened up and the 
pulsation was again detected (\cite{wilson1997}). Unfortunately, no PCA
coverage was possible for more than 2 months around this time because the Sun 
was too close to the source. Figure~1 summaries the X--ray flux history.
Pulse-period folded light curves were obtained from the Standard2 data with a 
16 s time resolution. From the folded light curves, we then computed the 
average pulse fraction, which is defined as the ratio of the difference 
between the maximum and minimum count rates to the maximum rate. The results
are also shown in Fig.~1. Note that no pulsation is detected above 3$\sigma$
in the faint state, so for each observation the light curve is folded around 
a period that maximizes the residual.

An extensive monitoring campaign on GRO~J1744-28 started shortly after its 
discovery (\cite{giles1996}). At the beginning, GRO~J1744-28 was observed 
weekly. It appeared as a $\sim$2 Crab persistent source with giant type~II 
bursts (at more than 10 Crab). The sampling rate increased quickly to almost 
once per day when the source became fainter toward the end of April. By 1996
May 12, GRO~J1744-28 entered a period when the X--ray pulsation was only 
detected intermittently with significance no more than $6\sigma$. The 
persistent emission flux was about 50 mCrab. Such a faint state lasted for 
about 7 weeks. 
For this study, we have selected 13 observations to cover the entire period. 
In the same way as in Figure~1, the measured X--ray flux and average pulse 
fraction are shown in Figure~2 for GRO~J1744-28 (with the detected X--ray 
bursts removed). Here,
the pulse fraction was derived from the Binned-mode data with a 16-ms time 
resolution, except for the last observation where the Binned-mode data were
not available and the Event-mode data with higher time resolution were used
instead.

\subsection{Spectral Characteristics}
To derive the total X--ray flux, we modeled the observed spectra with a 
cut-off power law, which is typical of accreting X--ray pulsars 
(\cite{white1983}). For both sources, a 
Gaussian component is needed to mimic an iron $K_{\alpha}$ line at $\sim$6.4 
keV. Such a model fits the observed X--ray spectra of GX~1+4 reasonably well,
with reduced $\chi^2$ values in the range of 1-3, except for the first two
observations, for which the object was relatively bright. Similarly, for 
GRO~J1744-28, the 
spectra for the faint state can be fit fairly well by this simple model, 
although none of the results are formally acceptable in terms of reduced 
$\chi^2$ values for the data for which the source was relatively bright. By 
examining the 
residuals, we found that the significant deviation mostly resides at the very 
low energy end of the PCA range, where the PCA calibration is less certain. 
Therefore, the flux 
determination should not be very sensitive to such deviation for these hard 
X--ray sources. Note that we did not follow the usual practice of adding 
systematic uncertainties to the data in the spectral analysis, because it is 
still not clear how to correctly estimate the magnitude of those uncertainties
at different 
energies. The uncertainty in the flux measurement (as shown in Figures 1 and 
2) was derived by comparing the results from different PCUs. 

The observed spectrum varies significantly for GX~1+4. The photon index 
varies in the range of 0.4-1.7, and infered hydrogen column density, in
the range of 4 to 15$\times 10^{22}\mbox{ }cm^{-2}$, although the latter is 
poorly constrained in the faint state. The spectral cutoff occurs at 6-17 keV,
with an e-folding energy in the range 11-55 keV. As GX~1+4 approaches 
the faint state, the spectrum becomes harder; eventually the spectral 
cutoff becomes unmeasurable. An opposite trend is observed for GRO~J1744-28;
its spectrum changes only mildly. When it was relatively bright, the photon 
index is steady (1.3-1.4) and the column density is in the range 
3-5$\times 10^{22}\mbox{ }cm^{-2}$. The spectrum is cut off at 16-18 
keV, with an e-folding energy in the range 11-19 keV. However, as it
enters the faint state, the spectrum turns significantly softer: the photon 
index varies from 1.5 to 2.5 and the cutoff energy moves down to 5-8 keV 
with the e-folding energy in the range 15-26 keV. It is interesting to 
note that the harder spectrum completely recovers when the source brightens 
up again (as typified by the last observation in Fig.~2).

\subsection{Magnetic Field}
The absence of the X--ray pulsation when the pulsars were in a low brightness
state is a clear indication of propeller effects. For pulsars, the 
co-rotation radius, $r_{co}$, is derived by equating the Keplerian velocity 
to the co-rotating Keplerian velocity, i.e., $\Omega r_{co}=(GM/r_{co})^{1/2}$,
where $\Omega$ is the angular velocity of the neutron star, and M is the mass.
Therefore,
\begin{equation}
r_{co} = 1.7\times 10^8\mbox{ }P^{2/3} \left(\frac{M}{1.4M_{\odot}}\right)^{1/3}\mbox{ }cm
\end{equation}
where P is the neutron star spin period.

In the presence of an accretion disk, it is still uncertain how to determine
the outer boundary of the magnetosphere in a pulsar (e.g., \cite{ghosh1979};
\cite{arons1993}; \cite{ostriker1995}; \cite{wang1996}. To a good 
approximation, here we define the magnetospheric radius, $r_m$, as where the 
magnetic pressure balances the ram pressure of a spherically accretion flow. 
Assuming a dipole field at large distance from the neutron star, we have 
(\cite{lamb1973})
\begin{equation}
r_m = 2.7\times 10^8\mbox{ } \left(\frac{L_x}{10^{37}\mbox{ }erg\mbox{ }s^{-1}}\right)^{-2/7} \left(\frac{M}{1.4M_{\odot}}\right)^{1/7} \left(\frac{B}{10^{12}\mbox{ }G}\right)^{4/7} \left(\frac{R}{10^6\mbox{ }cm}\right)^{10/7}\mbox{ }cm,
\end{equation}
where $L_x$ is the bolometric X-ray luminosity, B is the surface polar 
magnetic field strength, and R is the neutron star radius.

The mass accretion onto the pulsar magnetic poles ceases when $r_{co}=r_m$. 
From equations (1) and (2), the magnetic field strength is given by
\begin{equation}
B=4.8\times 10^{10}\mbox{ }P^{7/6} \left(\frac{F_x}{1.0\times 10^{-9}\mbox{ }erg\mbox{ }cm^{-2}\mbox{ }s^{-1}}\right)^{1/2} \left(\frac{d}{1\mbox{ }kpc}\right) \left(\frac{M}{1.4M_{\odot}}\right)^{1/3} \left(\frac{R}{10^6\mbox{ }cm}\right)^{-5/2}\mbox{ }G 
\end{equation}
where $F_x$ is the minimum bolometric X-ray flux at which X--ray pulsation is
still detectable, and d is the distance to the source. 

For GX~1+4 and GRO~J1744-28, the observed 2-60 keV X-ray fluxes are $\sim$0.16 and $2.34 \times 10^{-9}\mbox{ }erg\mbox{ }cm^{-2}\mbox{ }s^{-1}$, 
respectively, when the X--ray pulsation becomes unmeasurable.
Correction to the measured 2-60 keV flux at both high and low energies needs
to be made in order to derive the bolometric X--ray flux. Because of the 
spectral cutoff, little flux is expected from above 60 keV for GRO~J1744-28. At
the low energy end ($<$ 2 keV), extending the power law (with a photon index
of about 1.5) does not add much flux either ($<$ 18\%). The correction for 
absorption yields less than 30\%. Therefore, the uncertainty lies primarily 
in the distance measurement which can only vary by a factor of 2 
(\cite{giles1996}). Assuming a distance of 8 kpc (\cite{giles1996}), 
$B\simeq 2.4\times 10^{11}$~G.
Similarly, for GX~1+4, the bolometric correction below 2 keV is small. At
high energies, no spectral cutoff is observed in the faint state, so it must 
be beyond the PCA passing band. At the beginning of the faint state, the 
observed photon index is $\sim$1.7, so it is highly unlikely that the 
bolometric flux is more than a factor of 2 higher than what is measured 
(requiring a cutoff energy of $\sim$300 keV). However, the spectrum seems to 
harden in the faint state, and the photon index can drop as low as 0.5.
Even so, a spectral cutoff at $\sim$100 keV would be required for a 100\%
bolometric correction above 60 keV. Assuming a distance of 6 kpc (which can
vary by no more than a factor of 2; \cite{chakrabarty1997a}) and using the 
measured 2-60 keV flux in equation~(3), we have $B\gtrsim 3.1\times 10^{13}$
G for GX~1+4. In conclusion, the derived magnetic-field values are likely
to be accurate to within a factor of two for both sources, and the 
total uncertainty is dominated by the uncertainty in estimating the distances.

\section{Discussion}

What is the origin of the persistent emission in the faint state? The 
different (nearly opposite) spectral characteristics strongly suggest
different emission mechanisms for GX~1+4 and GRO~J1744-28. GX~1+4 has an 
M-giant companion star, and a relatively dense, slow stellar wind is expected 
in the system(\cite{chakrabarty1997a}). In fact, the comparable spin-up and 
spin-down rates observed seems to suggest the presence of a retrograde 
accretion disk during the spin-down period (\cite{chakrabarty1997b}), which is
only possible in a wind-fed system (as opposed to the Roche lobe overflow 
system). Because GX~1+4 has a strong magnetic field, 
accreting matter may not be able to penetrate the field lines very much and 
is likely to be flung away at a very high velocity. The velocity can be 
roughly estimated as $v \simeq (2GM/r_{co})^{1/2}$ (\cite{illarionov1975}), 
which is $\sim$2000 km/s. As the material plows through the dense stellar wind
at such a high velocity, a shock is bound to form. The observed emission in 
the faint state may simply be the synchrotron radiation by relativistic 
particles accelerated by the shock through mechanisms like Fermi acceleration 
(\cite{fermi1949}). The nonthermal nature of such emission is consistent with 
the disappearance of the spectral cutoff in the faint state. This emission 
mechanism is thought to be responsible for the unpulsed X--ray emission 
observed from the Be binary pulsar system PSR~B1259-63 near periastron 
(\cite{grove1995}).

For GRO~J1744-28, the magnetic field is very weak for an X--ray pulsar. When 
the propeller effects take place, a significant amount of accreting material 
might leak through ``between the field lines'' and reach the neutron star 
surface (\cite{arons1980}). The observed emission in the faint state would 
then come from a large portion of the surface and so would not be pulsed. The 
surface temperature reached is expected to be lower than that of the hot 
spots; thus the spectrum is softer, which is consistent with the observation. 

However, both sources are in the Galactic center region, so source confusion 
could be a serious problem. It is natural to question whether we actually 
detected the sources when the X--ray pulsation was not seen. We have 
carefully searched the catalogs for known X-ray sources within a 
$1^{\circ}$ radius circle around each source. None are found around GX~1+4. 
Moreover, the iron line at $\sim$6.4 keV is particularly prominent in GX~1+4
(see Fig.~3). Its presence and distinct shape with respect to the continuum 
greatly boosted our confidence about detection of the object in the faint 
state.  GRO~J1744-28 was still fairly bright ($\sim$50 mCrab) even in the 
faint 
state. Eighteen known X--ray sources appeared within the search circle, six
of which can be brighter than 30 mCrab, but only GS~1741.2-2859 and A~1742-289
are potentially close enough to the PCA pointing direction ($<$ 30\arcmin)
and bright enough to contribute significantly to the observed counts. 
However, both are X--ray transients, and are currently off, according to the 
nearly continuous monitoring with the All-Sky Monitor (ASM) aboard {\it RXTE}.
An extensive search for potential new 
sources in the region was also made with the ASM, but yielded null detection. 
The ASM long-term light curve indicates that GRO~J1744-28 was on throughout 
the faint state. In fact, type~II X--ray bursts were detected again by the
PCA during the later part of this period.

We seem to have selected two X--ray pulsars with their magnetic properties at 
opposite extremes. Both experience a period when the mass accretion from the 
companion star is hindered by the centrifugal barrier. The inferred magnetic 
field strength is in line with our previous knowledge about both sources. 
These results are important in constraining the details of the theoretical 
models on the X--ray emission mechanisms and thus will help us complete the
pictures on these sources, especially GRO~J1744-28, an unexpected bursting 
X--ray pulsar. It should be noted that the derived dipole field strength for
GRO~J1744-28 is consistent with the reported upper limits (\cite{finger1996}; 
 \cite{daumerie1996}; \cite{bildsten1997}) but is larger than the value derived
from the observed 40 Hz QPO (\cite{sturner1996}), based on the assumption that 
the ``beat frequency model'' of Alpar \& Shaham (1985) applies. This model is,
however, inconsistent with the observed dependence of the QPO properties on 
the X-ray brightness of the source (\cite{zhang1996}).

\acknowledgments
I would like to thank D.~Chakrabarty, A.~M.~Levine, S.~N.~Zhang, and J.~Li
for useful discussions. I have made use of data obtained through the High 
Energy Astrophysics Science Archive Research Center Online Service, provided 
by the NASA/Goddard Space Flight Center. This work is supported in part by 
NASA Contracts NAS5-30612.

\clearpage
\begin{figure}[t]
\epsfxsize=350pt \epsfbox{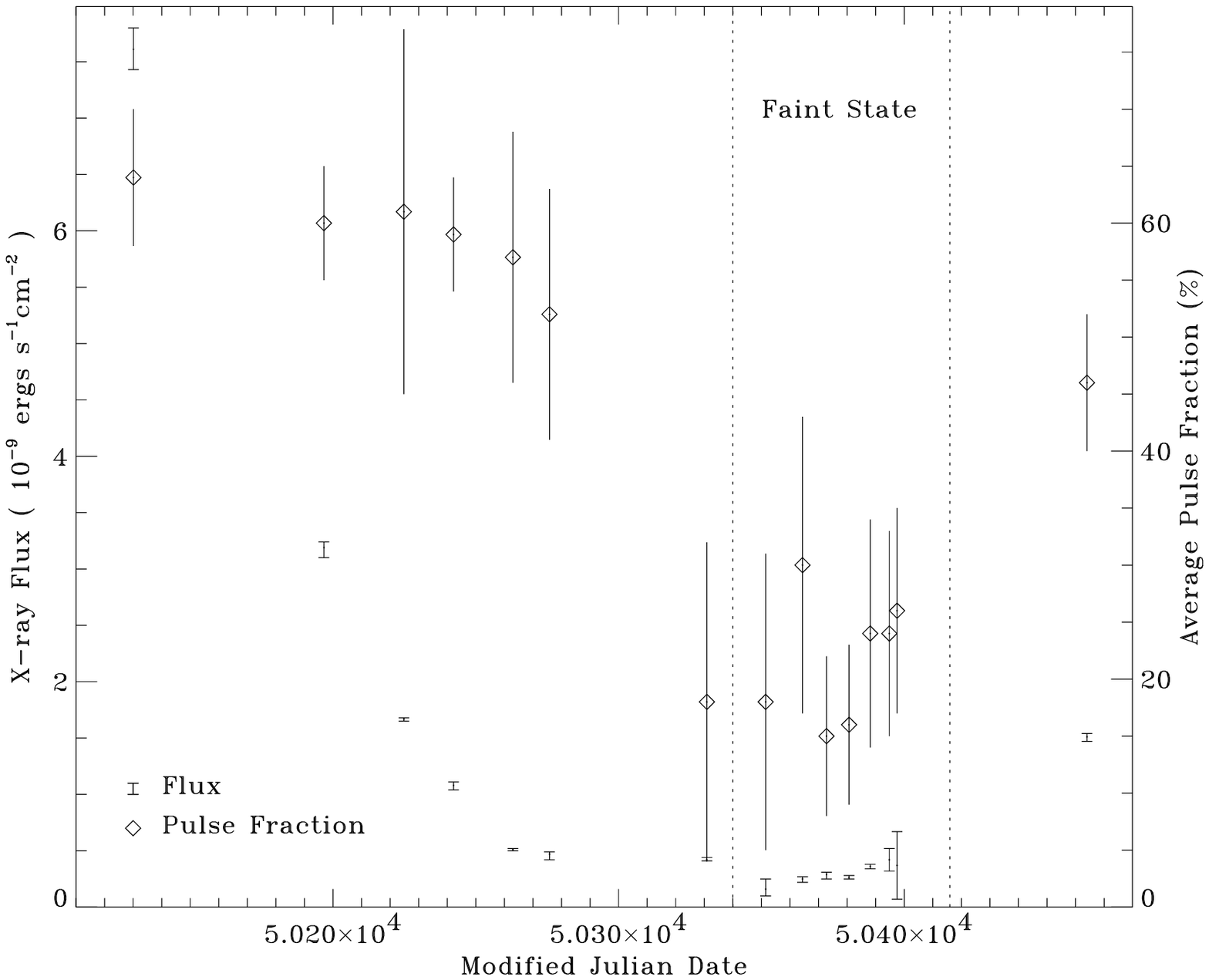}
\caption{X-ray flux and pulse fraction of GX~1+4. The fluxes were derived in 
the 2-60 keV band. Upper and lower limits are shown, representing the range 
of the flux measurements from different PCUs. See text for a definition of 
the pulse fraction. The error bars on the pulse fraction represent mean 
standard deviation. The time period between two dotted lines indicates the 
faint state when the X--ray pulsation was not detected. The left bound is 
only meant to indicate that it started some time between the two closest PCA 
observations. The right bound is from the BATSE monitoring of the source 
(Wilson \& Chakrabarty 1996). MJD 50351 corresponds to 25 September 1996. } 
\end{figure}

\clearpage
\begin{figure}[t]
\epsfxsize=350pt \epsfbox{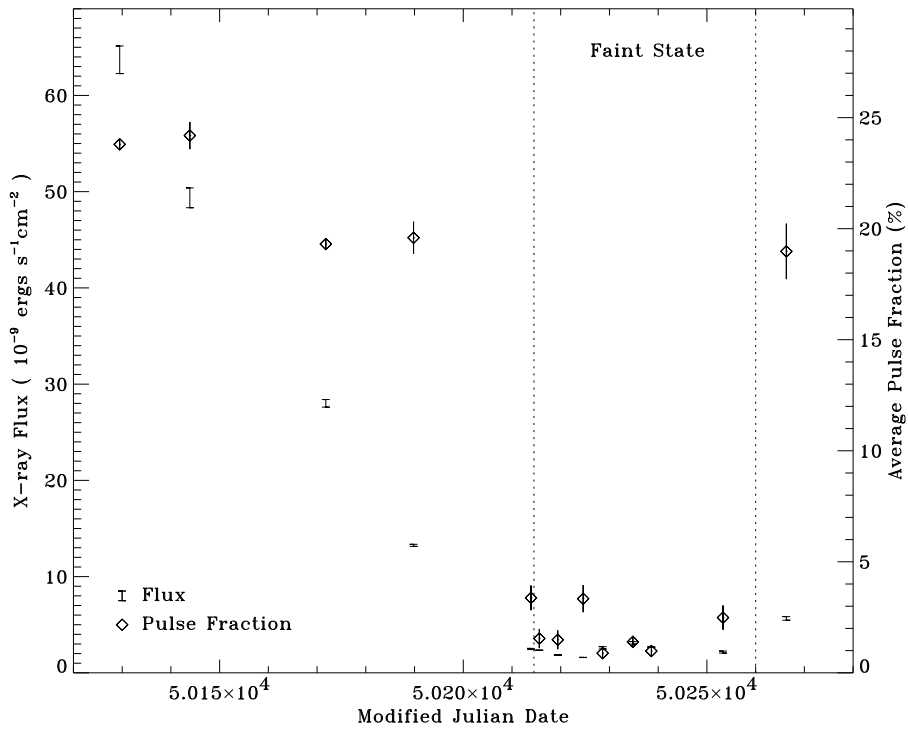}
\caption{The same as Figure~1, for GRO~J1744-28. In this case, the pulsation 
was only intermittently detected in the faint state. Again, the dashed lines 
are drawn arbitrarily between two adjacent PCA observations.}
\end{figure}

\clearpage
\begin{figure}[t]
\epsfxsize=350pt \epsfbox{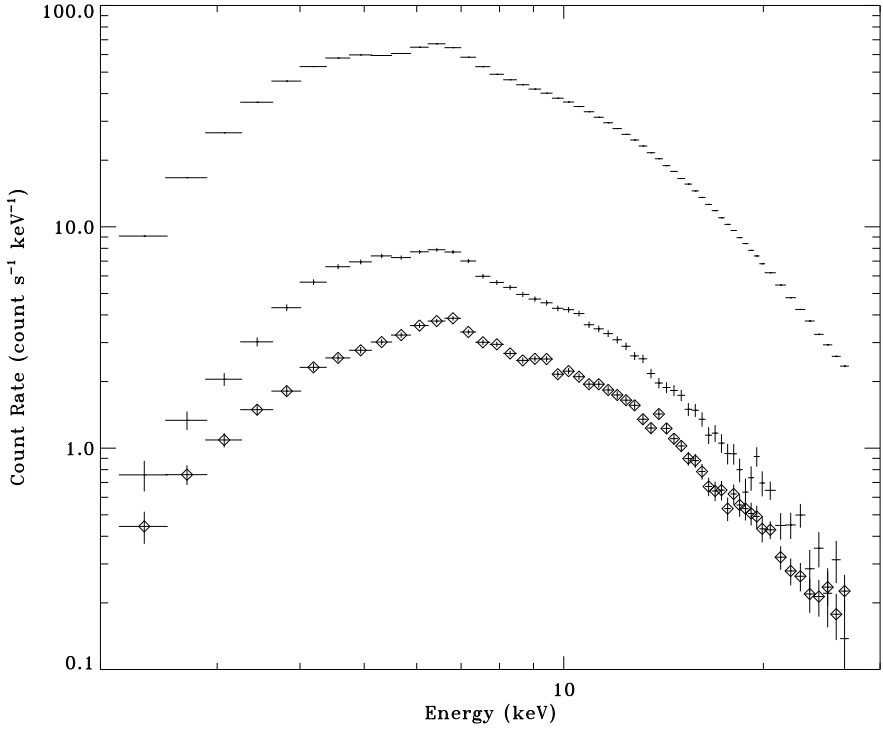}
\caption{Selected X--ray spectra of GX~1+4. The upper curve is 
derived from the 4/23/96 observation (MJD 50196.8), and is typical of the 
source when it is relatively bright. The middle curve is from the 7/11/96
observation (MJD 50275.9), and the bottom one from the 10/8/96 observation
(MJD 50364.4), which are meant to show typical X--ray spectra just before
and after the pulsation became undetectable, respectively. Note a strong
iron line $K_{\alpha}$ at $\sim$6.4 keV in all cases.}
\end{figure}

\clearpage

\begin{thebibliography}{}
\bibitem[Alpar \& Shaham 1985]{alpar1985}
Alpar,~M.~A., \& Shaham,~J. 1985, Nature, 316, 239
\bibitem[Arons 1993]{arons1993}
Arons,~J. 1993, \apj, 408, 160
\bibitem[Arons \& Lea 1980]{arons1980}
Arons,~J., \& Lea,~S.~M. 1980, \apj, 372, 565
\bibitem[Basko \& Sunyaev 1976]{basko1976}
Basko,~M.~M., \& Sunyaev,~R.~A. 1976, \aj, 20, 537
\bibitem[Bildsten \& Brown 1997]{bildsten1997}
Bildsten,~L, \& Brown,~E.~F. 1997, \apj, 477, 897
\bibitem[Chakrabarty, Finger, \& Prince 1996]{chakrabarty1996}
Chakrabarty,~D., Finger,~M.~H., \& Prince,~T.~A. 1996, \iaucirc\ 6478
\bibitem[Chakrabarty \& Roche 1997]{chakrabarty1997a}
Chakrabarty,~D., \& Roche,~P. 1997, \apj, submitted
\bibitem[Chakrabarty et al. 1997]{chakrabarty1997b}
Chakrabarty,~D., Bildsten,~L., Grunsfeld,~J.~M., Koh,~D.~T., Nelson,~R.,~W., Prince,~T.~A., \& Vaughan,~B.~A. 1997, \apjl, 481, in press
\bibitem[Cui \& Chakrabarty 1996]{cui1996}
Cui,~W., \& Chakrabarty,~D. 1996, \iaucirc\ 6478
\bibitem[Daumerie et al. 1996]{daumerie1996}
Daumerie,~P., Kalogera,~V., Lamb,~F.~K., \& Psaltis,~D. 1996, Nature, 382, 141
\bibitem[Fermi 1949]{fermi1949}
Fermi,~E. 1949, Phys. Rev. 75, 1169
\bibitem[Finger et al. 1996]{finger1996}
Finger,~M.~H., Koh,~D.~T., Nelson,~R.~W., Prince,~T.~A., Vaughan,~B.~A., \& Wilson,~R.~B. 1996, Nature, 381, 291
\bibitem[Ghosh \& Lamb 1979]{ghosh1979}
Ghosh,~P., \& Lamb,~F.~K. 1979, \apj, 234, 296
\bibitem[Giles et al. 1996]{giles1996}
Giles,~A.~B., et al. 1996, \apjl, 469, L25
\bibitem[Grove et al. 1995]{grove1995}
Grove,~J.~E., et al. 1995, \apjl, 447, L113
\bibitem[Illarionov \& Sunyaev 1975]{illarionov1975}
Illarionov,~A.~F., \& Sunyaev,~R.~A. 1975, \aap, 39, 185
\bibitem[Lamb, Pethick, \& Pines 1973]{lamb1973}
Lamb,~F.~K., Pethick,~C.~J., and Pines,~D. 1973, \apj, 184, 271
\bibitem[Lewin et al. 1996]{lewin1996}
Lewin,~W.,~H.~G., Rutledge,~R.~E., Kommers,~J.~M., van Paradijs,~J., \& Kouveliotou,~C. 1996, \apjl, 462, L39
\bibitem[Ostriker \& Shu 1995]{ostriker1995}
Ostriker,~E.~C., \& Shu,~F.~H. 1995, \apj, 477, 813
\bibitem[Pringle \& Rees 1972]{pringle1972}
Pringle,~R.~E., \& Rees,~M.~J. 1972, \aap, 21, 1
\bibitem[Sturner \& Dermer 1996]{sturner1996}
Sturner,~S.~J., and Dermer,~C.~D. 1996, \apjl, 465, L31
\bibitem[Wang 1996]{wang1996}
Wang,~Y.-M. 1996, \apj, 465, L111
\bibitem[White, Swank, \& Holt 1983]{white1983}
White,~N.~E., Swank,~J.~H., \& Holt,~S.~S. 1983, \apj, 270, 711
\bibitem[Wilson \& Chakrabarty 1997]{wilson1997}
Wilson,~R.~B., \& Chakrabarty,~D. 1997, \iaucirc\ 6536
\bibitem[Zhang et al. 1996]{zhang1996}
Zhang,~W., Morgan,~E., Swank,~J., Jahoda,~K., Jernigan,~G., \& Klein,~R. 1996, \apjl, 469, L29
\end{thebibliography}
\end{document}